\documentclass[prl,twocolumn,superscriptaddress]{revtex4-1}
\pdfoutput=1
\usepackage{listings}
\usepackage{natbib}
\usepackage{amsmath}    
\usepackage{pgf}   
\usepackage{hhline}
\usepackage{verbatim}   
\usepackage{color}      
\usepackage[caption=false]{subfig}

\usepackage{hyperref}   

\begin{document}

\title{Long-Lived Ultracold Molecules with Electric and Magnetic Dipole Moments}

\author{Timur M. Rvachov}
\affiliation{Research Laboratory of Electronics, MIT-Harvard Center for Ultracold Atoms,
Department of Physics, Massachusetts Institute of Technology, Cambridge, Massachusetts 02139, USA}

\author{Hyungmok Son}
\affiliation{Research Laboratory of Electronics, MIT-Harvard Center for Ultracold Atoms,
Department of Physics, Massachusetts Institute of Technology, Cambridge, Massachusetts 02139, USA}
\affiliation{Department of Physics, Harvard University, Cambridge, Massachusetts 02138, USA}

\author{Ariel T. Sommer}
\affiliation{Research Laboratory of Electronics, MIT-Harvard Center for Ultracold Atoms,
Department of Physics, Massachusetts Institute of Technology, Cambridge, Massachusetts 02139, USA}

\author{Sepehr Ebadi}

\affiliation{Research Laboratory of Electronics, MIT-Harvard Center for Ultracold Atoms,
Department of Physics, Massachusetts Institute of Technology, Cambridge, Massachusetts 02139, USA}
\affiliation{Department of Physics, University of Toronto, Toronto, Ontario M5S 1A7, Canada}

\author{Juliana J. Park}
\affiliation{Research Laboratory of Electronics, MIT-Harvard Center for Ultracold Atoms,
Department of Physics, Massachusetts Institute of Technology, Cambridge, Massachusetts 02139, USA}

\author{Martin W. Zwierlein}
\affiliation{Research Laboratory of Electronics, MIT-Harvard Center for Ultracold Atoms,
Department of Physics, Massachusetts Institute of Technology, Cambridge, Massachusetts 02139, USA}

\author{Wolfgang Ketterle}
\affiliation{Research Laboratory of Electronics, MIT-Harvard Center for Ultracold Atoms,
Department of Physics, Massachusetts Institute of Technology, Cambridge, Massachusetts 02139, USA}

\author{Alan O. Jamison}
\affiliation{Research Laboratory of Electronics, MIT-Harvard Center for Ultracold Atoms,
Department of Physics, Massachusetts Institute of Technology, Cambridge, Massachusetts 02139, USA}

\date{\today}
\pacs{123456790}

\begin{abstract}
We create fermionic dipolar $^{23}$Na$^6$Li molecules in their triplet ground state from an ultracold mixture of $^{23}$Na and $^6$Li. Using magneto-association across a narrow Feshbach resonance followed by a two-photon STIRAP transfer to the triplet ground state, we produce $3\,{\times}\,10^4$ ground state molecules in a spin-polarized state.  We observe a lifetime of $4.6\,\text{s}$ in an isolated molecular sample, approaching the $p$-wave universal rate limit. Electron spin resonance spectroscopy of the triplet state was used to determine the hyperfine structure of this previously unobserved molecular state.

\end{abstract}

\maketitle

Ultracold molecules with permanent electric dipole moments have gained considerable attention in recent years as promising new systems to study quantum chemistry and quantum many-body physics \cite{chemistry_review_krems_2008,review_ye_2009,dipole_review_lewenstein_2011,dipole_review_zoller_2012}.  An additional magnetic dipole moment provides an extra degree of control that can be used for magnetic trapping, tuning collisions and chemical reactions \cite{chemistry_review_krems_2008, magmoment_timur_2007}, simulation of spin-lattice Hamiltonians \cite{micheli_2006}, or as in the case of cold magnetic atoms, direct study of magnetic dipole interaction effects \cite{pfau_dipole_bec,dipole_review,pfau_nature}.

Experimental realization of ultracold molecules with magnetic and electric dipole moments is an ongoing challenge. Direct cooling methods have demonstrated promising advances including magnetic \cite{magtrap_doyle} and magneto-optical trapping (MOT) \cite{mot_demille_nature, mot_doyle}, however temperatures are thus far limited to the ${>}\,100\,\mu$K regime and at low densities (${<}\,10^7\,\text{cm}^{-3}$). To date, cold molecules nearest to quantum degeneracy have been achieved using a different approach: formation from an initial cloud of ultracold atoms  \cite{cs2_hcn_2008,krb_ni_2008}. This method has been widely successful, including KRb \cite{krb_ni_2008}, RbCs \cite{rbcs_hcn_2014,rbcs_cornish_2014}, NaK \cite{nak_park_2015}, and NaRb \cite{narb_wang_2016}. In heteronuclear cases, long-lived dipolar molecules have only been created in the singlet state, i.e. without an electronic magnetic dipole moment. Studies of dipolar triplet molecules were limited to lifetimes of only ${\sim}\,200\, \mu$s \cite{krb_ni_2008}. 

$^{23}$Na$^6$Li is a dipolar fermionic molecule that is particularly well-suited to study the triplet ground state. It has weak singlet-triplet mixing and weak two-body scattering rates as predicted by the universal model for cold collisions, allowing for long lifetimes despite possible loss channels such as chemical reactivity \cite{univrates_julienne_2011, reactive_hutson_2010}. The universal rate constant for inelastic loss in NaLi is the smallest of the heteronuclear bi-alkalis, a factor of 10 lower than in KRb \cite{univrates_julienne_2011,krb_ospelkaus_2010}. Triplet NaLi has a magnetic dipole moment of $2\mu_B$ arising from the aligned electron spins, and a predicted electric dipole moment of $0.2\,\text{D}$, which is one of the largest among the triplet states of bi-alkalis \cite{dipole_moments_dulieu,dmoment_krems_2013}.

\begin{figure}[h]
\includegraphics[width=0.5\textwidth]{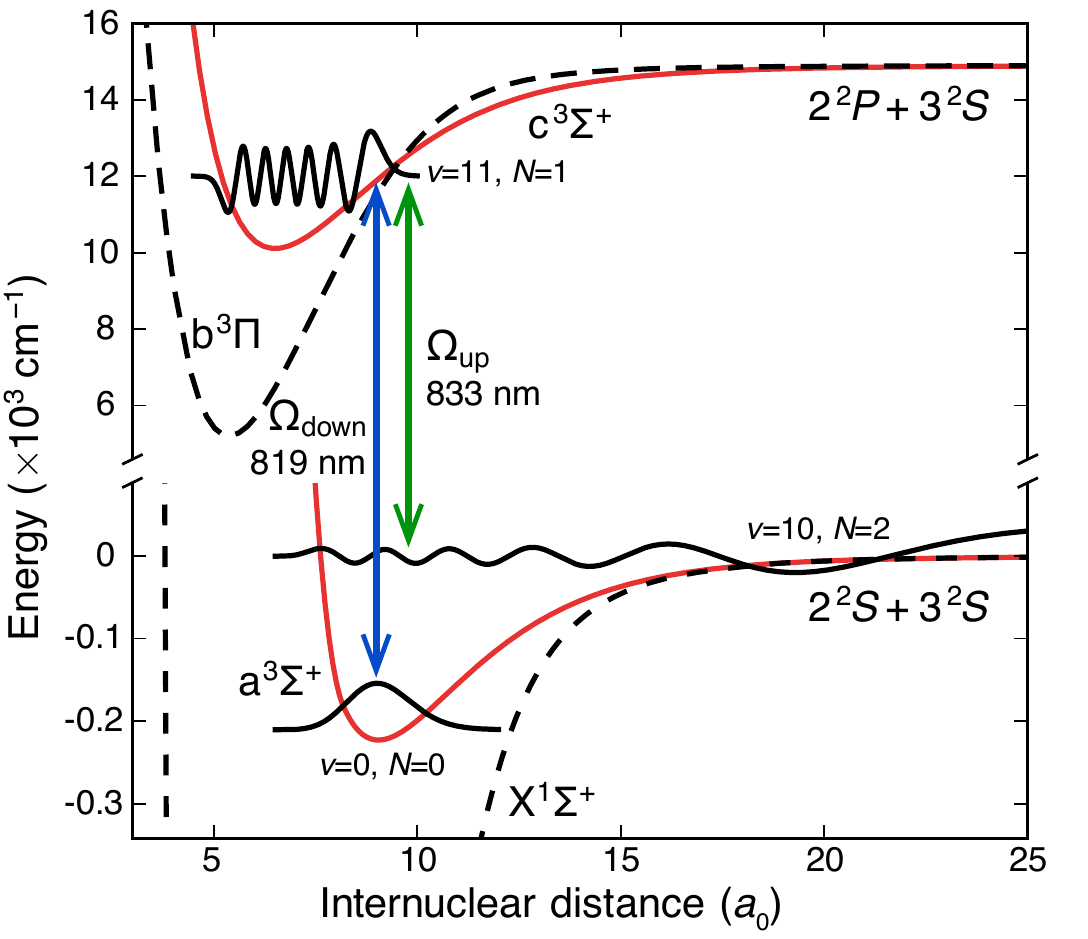}\\[-3ex]
\caption{Molecular potential diagram for $^{23}$Na$^6$Li. The potential energy curves relevant to STIRAP (solid red) and other nearby potentials (dashed black) are shown.  The radial wavefunctions are drawn for the Feshbach molecular state, the intermediate excited state, and the triplet ground state.}
\label{fig:molpot}
\end{figure}

In this Letter, we realize long-lived samples of NaLi triplet ground state molecules with a lifetime of $4.6\,\text{s}$, a $10^4$ fold improvement on previous triplet lifetimes in ultracold dipolar molecules \cite{krb_ni_2008}.  The molecules are formed using magneto-association around a Feshbach resonance, followed by a two-photon stimulated Raman adiabatic passage (STIRAP) to the ro-vibrational ground state (Fig.~\ref{fig:molpot}) \cite{krb_ni_2008,cs2_hcn_2008}.  We probe the molecular hyperfine structure using electron spin resonance (ESR) spectroscopy, a technique only feasible in molecules with a magnetic moment \cite{krb_hyperfine_2010}. In this way, we measure the hyperfine constants of this previously unobserved molecular state \cite{nali_singlet_tiemann}.

\begin{figure}[!ht]
\subfloat{%
  \includegraphics[clip,width=\columnwidth]{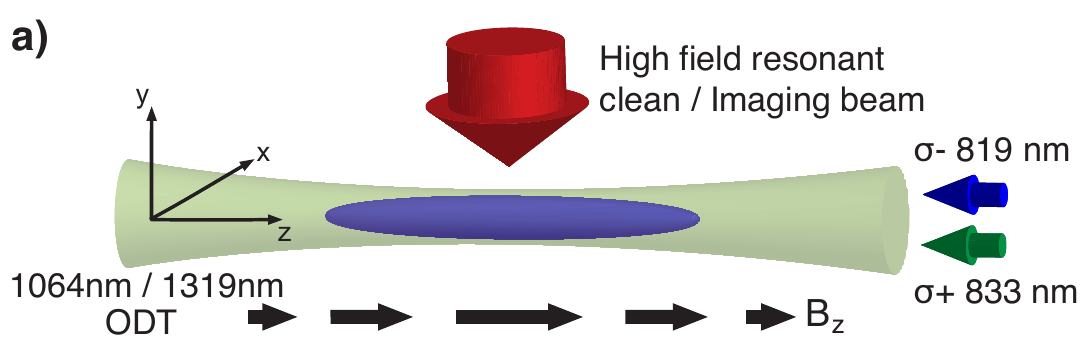}%
}

\subfloat{%
  \includegraphics[clip,width=\columnwidth]{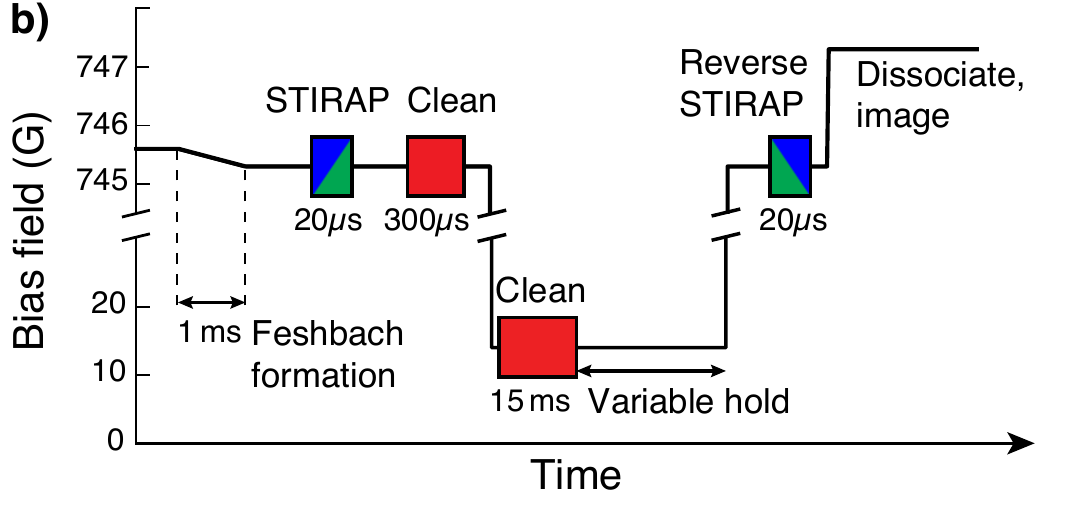}
 }\\[-3ex]
\caption{Experimental setup. a) Trap and laser geometry. The low field MOT cleanout beams are not shown. b) Experimental sequence to produce and isolate NaLi molecules (time axis is not to scale). Molecules are formed via Feshbach formation and STIRAP, and the free atoms are removed using two resonant light pulses at low and high field. For detection, the molecule formation process is reversed, and the dissociated free atoms are imaged. Experiments at short (${<}\,20\,\text{ms}$) hold times were performed at high field, without low-field cleanout.}
\label{fig:sequence}
\end{figure}

We produce an ultracold mixture of $^{23}$Na and $^6$Li in their lowest energy Zeeman sublevels (which adiabatically connect to Na $|F,m_F\rangle\,{=}\,|1,1\rangle$ and Li $|1/2,1/2\rangle$ at low field), at a bias field of $B_\text{z}\,{=}\,745.55\,\text{G}$ which is $150\,\text{mG}$ above an interspecies Feshbach resonance \cite{nali_feshbach_original,feshspec_tiemann_2012}. The mixture is confined to a cigar-shaped potential from a 1064\,nm optical dipole trap (ODT), ($0.3$ to $2\,\text{W}$, $30\,\mu \text{m}$ waist) coaxial to the bias field. After $2.5\,\text{s}$ of forced evaporation, we transfer the sample to a coaxial $1319\,\text{nm}$ ODT ($250\,\text{mW}$, $32\,\mu\text{m}$ waist) in order to suppress photon scattering in molecular transitions discussed below. The final mixture contains $1.5\,{\times}\, 10^6$ Na and $1\,{\times}\, 10^6$ Li atoms at a temperature of $2\,\mu\text{K}$, which corresponds to peak densities and temperatures of  $n_\text{Na}\,{\approx}\,4\,{\times}\,10^{12}\,\text{cm}^{-3}$, $T/T_c\,{\approx}\,3$ and $n_\text{Li}\,{\approx}\,3\,{\times}\,10^{12}\,\text{cm}^{-3}$, $T/T_F\,{\approx}\,1$, where $T_c$, $T_F$ are the in-trap condensation and Fermi temperatures, respectively. The radial confinement is produced by the ODT,  with trap frequencies $f_{r,\text{Na}}\,{=}\,0.5\,\text{kHz}$, $f_{r,\text{Li}}\,{=}\,1\,\text{kHz}$. Axial confinement is produced from both the ODT and the curvature associated with the Feshbach bias coils, with $f_{z,\text{Na}}\,{=}\,13\,\text{Hz}$ and $f_{z,\text{Li}}\,{=}\,26\,\text{Hz}$ (Fig.~\ref{fig:sequence}a).

\mbox{Formation} of Feshbach molecules around the narrow 745 G resonance has been previously demonstrated \cite{feshmol_ketterle_2012}, and we briefly discuss the procedure here. The 745\,G Feshbach bias field is actively stabilized to ${<}\,3\,\text{mG}_\text{rms}$ noise in the  $1\,\text{Hz}\,{-}\,10\,\text{kHz}$ band, and Feshbach molecule formation sweeps are performed with a small shim coil to reduce eddy current field delays.  The field is lowered by $290\,\text{mG}$ in $1\,\text{ms}$ across the resonance to form $3\,{\times}\,10^4$ molecules in the $a^3\Sigma^+$, $\nu\,{=}\,10$, $N\,{=}\,2$, $m_N\,{=}\,{-}2$ state \cite{feshspec_tiemann_2012,feshmol_ketterle_2012}, corresponding to a formation efficiency of 3\% ($\nu$, $N$ are the vibrational and rotational molecular quantum numbers, respectively). The Feshbach molecules are isolated using a $300\,\mu\text{s}$ light pulse resonant with the free atoms, without affecting the molecules. The pulse has intensities of $I_\text{Na}\,{=}\,80\,\mu\text{W}/\text{cm}^2$, $I_\text{Li}\,{=}\,15\,\mu\text{W}/\text{cm}^2$ (Fig.~\ref{fig:sequence}a). The cleanout pulse leaves ${<}\,2000$ Li atoms, however $5\,{\times}\,10^4$ Na atoms are left in the trap, since they are pumped into the upper $F\,{=}\,2$ Zeeman manifold and become dark to the cleanout light. These residual Na atoms limit the collisional lifetime of the Feshbach molecules. Molecules are detected by jumping the field $2\,\text{G}$ above the Feshbach resonance, and imaging the dissociated molecules in absorption.

\begin{figure}[h]
\includegraphics[width=0.5\textwidth]{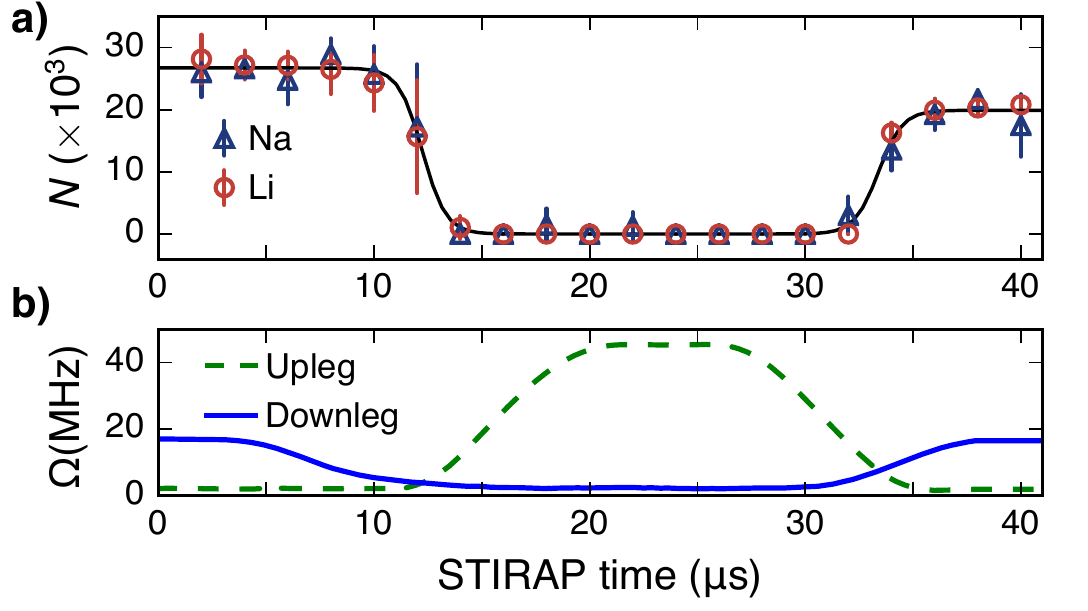}\\[-2ex]
\caption{STIRAP sequence. a) Number of Feshbach molecules monitored through free atom Na and Li numbers after a dissociation ramp, taken at various times across a forward and reverse STIRAP, demonstrating coherent transfer down to the ground state and back. b) The Rabi frequencies of the upleg and downleg transition during the STIRAP pulse sequence.}
\label{fig:stirap}
\end{figure}

Implementation of STIRAP required the characterization of $^3\Sigma$ excited and ground states via one- and two-photon spectroscopy \cite{nali_singlet_tiemann,in_prep_footnote,ourtwophoton}.  Transfer from the Feshbach state to the triplet ground state ($a^3\Sigma^+$, $\nu\,{=}\,0$, $N\,{=}\,0$) was performed using two-photon STIRAP via the intermediate state in the excited state potential, $c^3\Sigma^+$ ($\nu\,{=}\,11$, $N\,{=}\,1$, $m_N\,{=}\,{-}1$) which can be reached with an ``upleg'' laser at 833 nm ($360.00381(1)\,\text{THz}$) and a ``downleg'' laser at 819 nm ($366.23611(1)\,\text{THz}$) (Fig.~\ref{fig:molpot}). The nuclear spin states ($m_{I,\text{Na}}\,{=}\,3/2$, $m_{I,\text{Li}}\,{=}\,1$) are unaffected by the STIRAP.  The intermediate state was chosen to maximize the downleg Franck-Condon factor. To produce the narrow relative linewidth laser light necessary for STIRAP, we lock a Ti:Sapphire laser for the upleg and a homebuilt external cavity diode laser \cite{steck_ecdl} for the downleg to an ultra-low-expansion cavity, achieving a relative linewidth between the two lasers of ${<}\,1\,\text{kHz}$. The STIRAP beam is coaxial with the bias field with $\sigma^+$ polarization for the upleg and $\sigma^-$ polarization for the downleg to provide the two quanta of rotational angular momentum difference between the Feshbach state and the ground state. The downleg Rabi frequency is $\Omega_\text{down}/\sqrt{I}\,{=}\,2\pi\,{\times}\,580\,\text{kHz}/(\sqrt{\text{mW}/\text{cm}^2})$, which was measured via two-photon spectroscopy of the Autler-Townes doublet. The upleg Rabi frequency is $\Omega_\text{up}/\sqrt{I}\,{=}\,2\pi\,{\times}\,28\,\text{kHz}/(\sqrt{\text{mW}/\text{cm}^2})$, which was determined by driving the upleg transition starting with the Feshbach molecules, and fitting the loss rate to an exponential decay with time constant $\tau^{-1}\,{=}\,\Omega_\text{up}^2/\Gamma$, where we have assumed the excited state has a natural linewidth of $\Gamma\,{=}\,2\pi\,{\times}\,6\,\text{MHz}$. Detection of ground state transfer is performed with a reverse-STIRAP pulse (Fig.~\ref{fig:stirap}) back into the Feshbach state, followed by dissociation to free atoms.  Optimal ground state transfer with one-way efficiency of 86\% is achieved with a $20\,\mu\text{s}$ STIRAP pulse.

In order to measure the lifetime of isolated NaLi molecules, the field is reduced to $14\,\text{G}$ after molecule formation (Fig.~\ref{fig:sequence}b). This is necessary for two reasons: First, the high-field cleanout light pumps a fraction of Na into a Zeeman dark state, and full removal of free atoms is achieved at low field with a $15\,\text{ms}$ MOT-light pulse. Second, the molecules are in a low field seeking state, which is anti-trapped at high fields due to the slight curvature produced by the Feshbach bias coils. At low field, the molecules are confined in all three dimensions by the optical trap.  Using a single beam $1064\,\text{nm}$ ODT resulted in molecule lifetimes of $200\,\text{ms}$, despite the thorough cleanout of free atoms at low field. Using ab-initio potentials as guidance, we suspect $1064\,\text{nm}$ light drives a bound-bound molecular transition to the $b^3\Pi$, $\nu\,{\sim}\,16$ excited state. Ab-initio potentials predict much lower scattering rates at longer wavelengths, and we find that forming molecules in a $1319\,\text{nm}$ ODT increased the lifetime by a factor of 30. 

\begin{figure}[!t]
\subfloat{%
  \includegraphics[clip,width=\columnwidth]{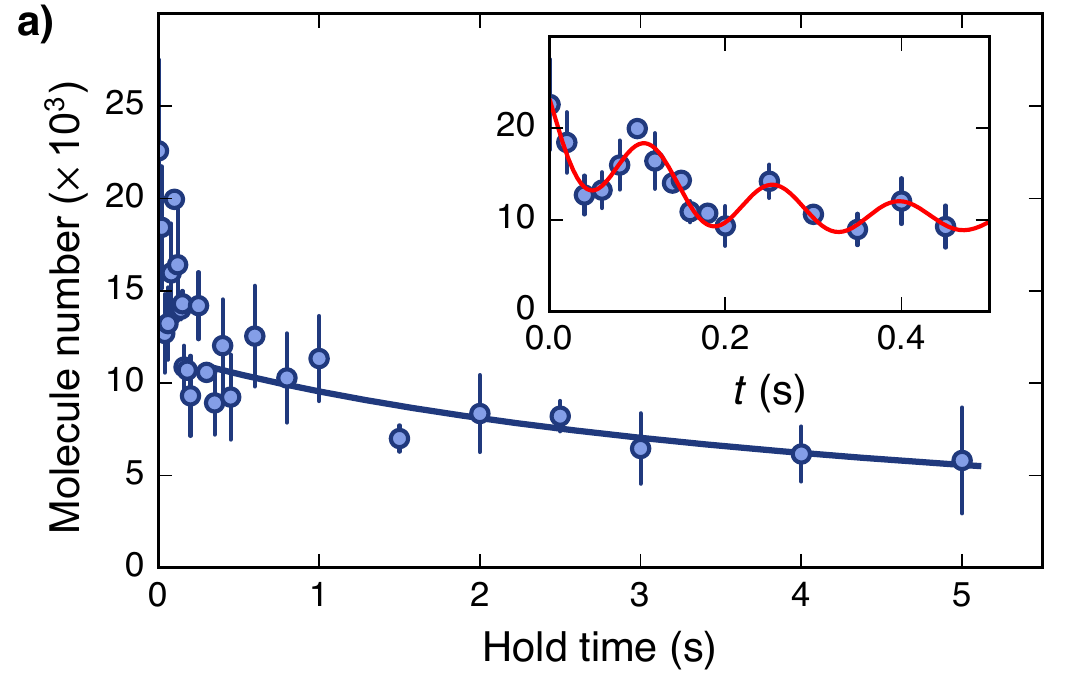}%
}\\[-1ex]

\subfloat{%
  \includegraphics[clip,width=\columnwidth]{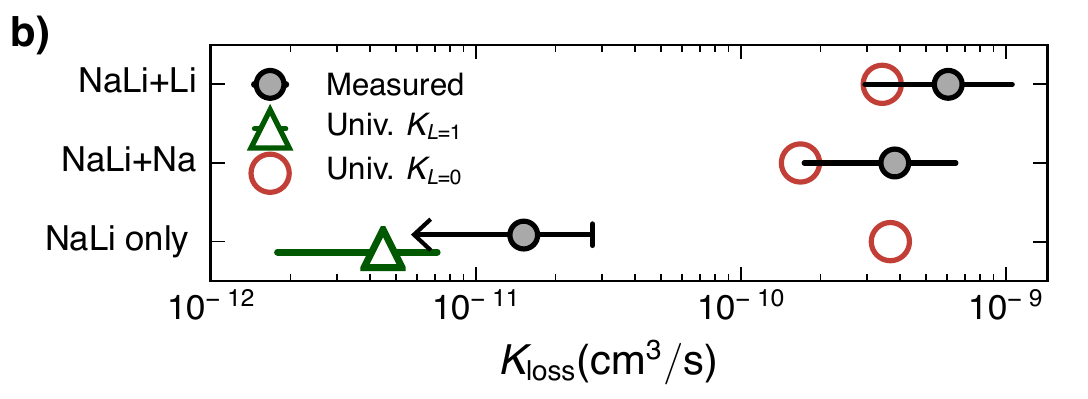}
}\\[-3ex]

\caption{Molecule lifetimes. a) Molecule number as a function of hold time at $14\,\text{G}$ bias field. After rapid initial decay due to a kick from the sudden magnetic field jump, the molecules are long-lived with a fitted two-body loss lifetime of $\tau \equiv (K n_\text{mol})^{-1}=4.6\,\text{s}$ for $t\,{>}\,300\,\text{ms}$. Inset: At short hold times a clear $7$ Hz oscillation is visible from the axial sloshing, which periodically modulates the detection efficiency.  b) Two-body loss coefficients for NaLi with Li, Na, and NaLi by itself.  The error bars arise from temperature uncertainty and the inhomogenous density distribution. The error bar on the ``NaLi only'' rate indicates our measurement is an upper bound.}
\label{fig:lifetime}
\end{figure}

The NaLi ground state lifetime is shown in Fig.~\ref{fig:lifetime}. We fit the lifetime data to a two-body loss model,
\begin{equation}
\dot N = -KN^2/V_\text{eff} \quad \rightarrow \quad N(t)=N_0/\left(1+t/\tau\right),
\end{equation}
where $N$ is the molecule number, $K$ is the density normalized loss coefficient, $V_\text{eff}$ is the effective volume, and $\tau\,{\equiv}\, N_0 K /V_\text{eff}$ is the characteristic two-body loss time scale. For long hold times ($t\,{>}\,300\,\text{ms}$), the fitted lifetime is $\tau\,{=}\,4.6\,\text{s}$. The effective volume \mbox{$V_\text{eff}=N^2/\int n^2 dV$} is given by the product of the three \mbox{Gaussian} waists in the molecule density distribution, i.e. \mbox{$V_\text{eff}=\prod_i 2\sqrt{\pi k_B T/m\omega_i^2}$}, where $T$ is the temperature, $m$ is the molecule mass, and $\omega_i$ are the trapping angular frequencies.  We determined the radial trapping frequency of the molecules by parametric modulation of the $1319\,\text{nm}$ ODT and the observation of a loss feature at $2f_{r,\text{mol}}\,{=}\,2\,{\times}\,770\,\text{Hz}$. Knowing the radial trapping frequency and the Gaussian beam geometry of the trapping laser, we determined the axial trap frequency $f_{z,\text{mol}}\,{=}\,7\,\text{Hz}$, and trap depth $U_\text{ODT, mol}\,{=}\, 20\, \mu\text{K}$. The axial frequency is consistent with the observed oscillation in the molecule lifetime (Fig.~\ref{fig:lifetime}a, inset) at short hold times.  The molecule temperature is estimated to be $T_\text{mol}\,{=}\,3^{+3}_{-1.5}\,\mu\text{K}$ by holding the molecules for 1 second (until the axial sloshing has settled) and measuring the temperature of the dissociated free atoms. The process of Feshbach dissociation can heat or cool the sample depending on the exact details of the dissociation ramp \cite{fesh_diss_ketterle_2004,feshdiss_rempe_2004}, and these effects are accounted for in the temperature uncertainty. The peak molecule density is calculated to be $n_\text{mol} \,{=}\, 5\,{\pm}\,4\,{\times}\, 10^{10} \,\text{cm}^{-3}$, with two-body loss coefficient $K_\text{meas}\,{=}\,1.6\,{\pm}\,1.2\,{\times}\,10^{-11}\,\text{cm}^{3}/\text{s}$.

Alongside lifetime measurements of an isolated NaLi molecule sample, we have characterized collisions of the ground state molecule with remaining Na $|1,1\rangle$ or Li $|1/2,1/2\rangle$ atoms by selectively only removing one of the species at high field, yielding lifetimes of $\tau\,{\approx}\,2\,\text{ms}$ (since these lifetimes were short, lowering the bias field for MOT-light cleanout was not necessary). We compare measured rates to the so-called universal rate which assumes unit probability loss at short range, therefore the rate is determined by long range (van der Waals) attraction. In the absence of resonances, the universal loss rate is an upper limit.  The universal two-body rates for $s$- and $p$-wave scattering channels are given by \cite{univrates_julienne_2011}: 
\begin{equation}
K_{L=0} = g\dfrac{4\pi \hbar }{\mu}\bar a \quad\quad K_{L=1}(T) = 1513\, \bar a^3 k_B T / h
\end{equation}
where $g\,{=}\,2$ for identical particles (otherwise, $g\,{=}\,1$), $\mu$ is the reduced mass of the collision partners, $\bar a \,{=}\, \left(2\pi/\Gamma(1/4)^2\right) (2\mu C_6/\hbar^2)^{1/4}$ is the van der Waals length, and $C_6$ is the van der Waals coefficient. The universal rates for molecule-atom collisions are given by $K_{L=0}$, while molecule-molecule collisions must occur through \mbox{$p$-wave} scattering due to their fermionic nature, hence we expect the universal limit of $K_{L=1}$.  The $C_6$ is estimated as the sum of dispersion coefficients for atomic pairs that constitute the collision partners \cite{lics_c6_weidemuller,li2_tout}. Fig.~\ref{fig:lifetime}b shows the loss coefficients obtained from our lifetime measurements. The fermionic nature of the NaLi molecule is highlighted by its low loss rate, $K_\text{meas}\,{\approx}\, 10^{-11}\, \text{cm}^3/\text{s}$, a factor of 20 lower than one would expect through the \mbox{$s$-wave} collision channel. The measured loss coefficient is still ${\sim}\,3$ times faster than the universal rate, $K_{L=1}$. Despite efforts to minimize single-photon scattering by using $1319\,\text{nm}$ light for optical trapping, we suspect there is still one-body loss due to the ODT driving transitions to the $b^3\Pi$ potential. Thus, the measured two-body loss coefficient is an upper bound.
\begin{figure}[!ht]
\subfloat{%
  \includegraphics[clip,width=\columnwidth]{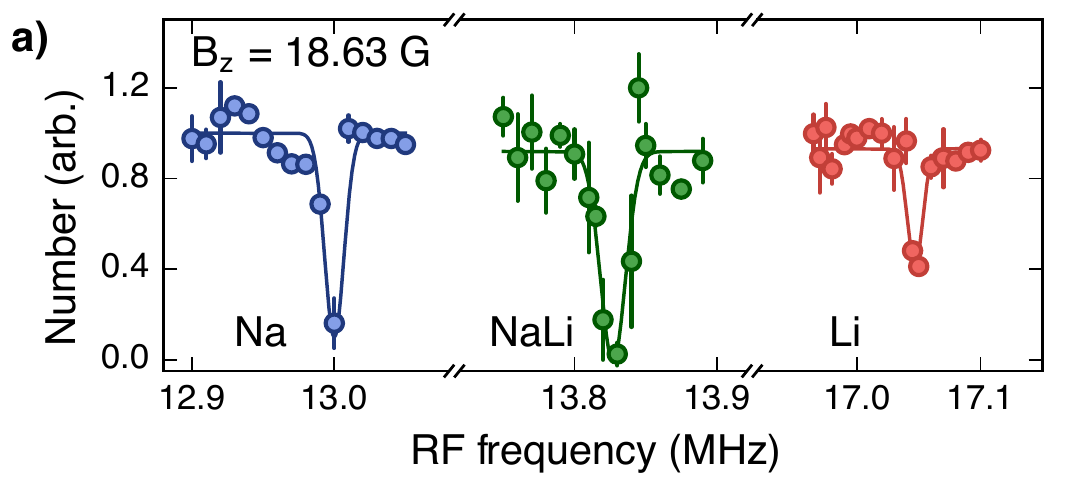}%
}\\[-1ex]

\subfloat{%
  \includegraphics[clip,width=\columnwidth]{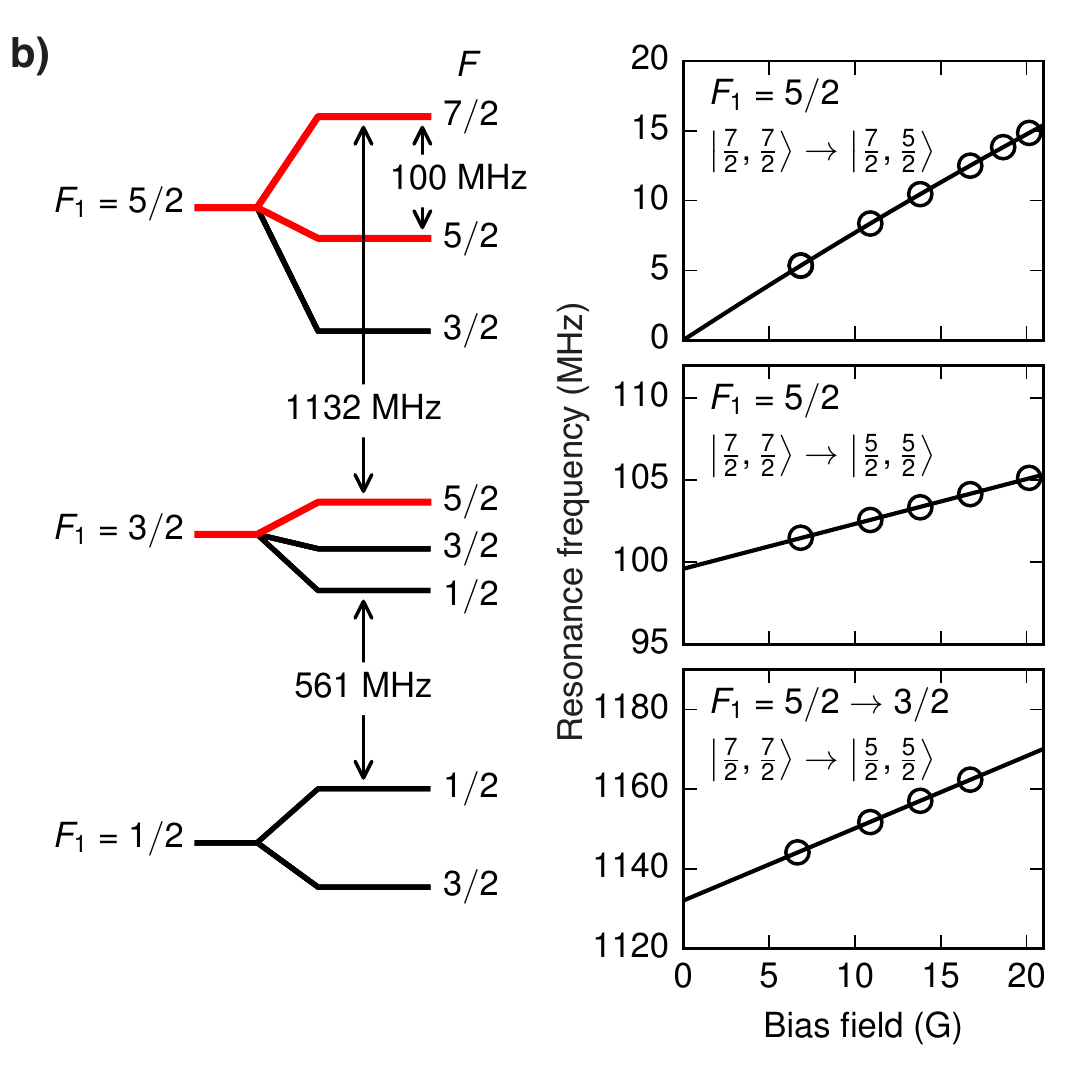}
 }\\[-3ex]
\caption{ESR spectroscopy. a) Sample spectra at $B_z\,{=}\,13.63\,\text{G}$ showing distinct loss features for transitions in Na $|F,m_F\rangle\,{=}\, |1,1\rangle \,{\rightarrow}\, |1,0\rangle$ (blue), NaLi $|7/2,7/2\rangle \,{\rightarrow}\, |7/2,5/2\rangle$ (green), and Li $|1/2,1/2\rangle\,{ \rightarrow}\, |1/2,-1/2\rangle$ (red). Typical interrogation times were $10\,{-}\,400\,\text{ms}$ with Rabi frequencies of $1\,{-}\,100\,\text{kHz}$. b) (Left) Energy level diagram of the NaLi hyperfine structure.  The three states accessible with a magnetic dipole transition from $|7/2,7/2\rangle$ must have $m_F=5/2$, which exist in the manifolds marked in red. (Right) The measured Zeeman shifts for each of the three transitions.}
\label{fig:rf}
\end{figure}

We demonstrate the utility of having non-zero electron spin in the triplet ground state by directly driving magnetic dipole spin flips of the aligned electrons \cite{krb_hyperfine_2010}. The hyperfine coupling Hamiltonian in $\nu\,{=}\,0$ is $H_\text{HF}=a_\text{Li} \vec S \cdot \vec I_\text{Li} + a_\text{Na} \vec S \cdot \vec I_\text{Na}$, where $a_\text{Na,Li}$ are the hyperfine coupling constants. Since the $a^3\Sigma^+$ potential is relatively shallow, we can use the Na, Li free atom hyperfine constants as guidance to understand the hyperfine structure of the molecules. The hyperfine coupling constant of $^{23}$Na is much larger than that of $^6$Li, i.e.  $a_\text{Na}\,{\gg}\, a_\text{Li}$, thus the electron spin first couples to $\vec I_\text{Na}$ to form $\vec F_1 \equiv \vec S + \vec I_\text{Na}$. This combined angular momentum couples to $\vec I_\text{Li}$ to form the total angular momentum $\vec F \equiv \vec F_1 + \vec I_\text{Li}$ (see Fig.~\ref{fig:rf}). We have performed ESR spectroscopy of the molecules at low bias-fields ($B_z \,{=}\, 4\,{-}\,20\,\text{G}$) and found all three accessible radio frequency (RF) transitions from our initial state, $|F,m_F\rangle\, {=}\,|7/2, 7/2 \rangle$. The low field ESR spectrum uniquely identifies the three species (Na, Li, NaLi) in our experiment (Fig.~\ref{fig:rf}a).  From the spectra we determined the hyperfine constants to be $a_\text{Li}=74.61(3)\,\text{MHz}$, $a_\text{Na}=433.20(3)\,\text{MHz}$. The hyperfine constants are approximately half of their corresponding free atom values, as observed in the $a^3\Sigma^+$ states of NaK \cite{nak_hyperfine}. This is expected from the shallow nature of the $a^3\Sigma^+$ potential: The wavefunctions of the two valence electrons are still localized to their respective atoms, but the total electronic spin is doubled.

In conclusion, we have demonstrated the formation of $3\,{\times}\,10^4$ ultracold $^{23}$Na$^6$Li molecules in their triplet ground state, with a 4.6 second lifetime at peak densities of $n_\text{mol} = 5(4) \,{\times}\, 10^{10}\,\text{cm}^{-3}$. We have used ESR to measure the ground state hyperfine structure. Triplet NaLi is the lightest heteronuclear bi-alkali, and could provide insight into cold molecule collisions where recent experiments \cite{rbcs_hcn_2014,rbcs_cornish_2014,nak_park_2015,narb_wang_2016}  indicate the possibility of novel two-body loss processes that scale with molecular mass \cite{sticky_bohn_2013}. In the near future, triplet NaLi molecules will enable the use of magnetic trapping and  collisional studies between fully spin-polarized atomic collision partners, perhaps allowing for sympathetic cooling towards quantum degeneracy \cite{reactivity_quartet,dmoment_krems_2013}. 
\begin{acknowledgments}
We would like to thank Travis Nicholson, Jee Woo Park, and Sebastian Will for valuable technical discussions, as well as William Cody Burton and Lawrence Cheuk for critical reading of the manuscript. We acknowledge support from the NSF through the Center for Ultracold Atoms and award 1506369, from the ONR DURIP grant N000141613141, and from MURIs on Ultracold Molecules (AFOSR grant FA9550-09-1-0588) and on Quantized Chemical Reactions of Ultracold Molecules (ARO grant W911NF-12-1-0476). H.S. and J.J.P. acknowledge additional support from Samsung Scholarship. 
\end{acknowledgments}

\bibliographystyle{apsrev4-1}

%

\end{document}